\documentclass[a4paper,11pt]{article}
\usepackage{pos}

\usepackage{physics}
\usepackage{float}
\usepackage{amsmath}
\usepackage{hyperref}
\usepackage{comment}
\usepackage{subcaption,siunitx,booktabs}

\title{Predicting the spectrum and decay constants of positive-parity heavy-strange mesons using domain-wall fermions}
 \ShortTitle{Spectrum and decay constants of positive-parity heavy-strange mesons}

\author*{Forrest Guyton}
\author{Stefan Meinel}

\affiliation{Department of Physics, University of Arizona, 
  Tucson, AZ 85721, USA}
\emailAdd{forrestguyton@arizona.edu}
\emailAdd{smeinel@arizona.edu}

\abstract{We present a lattice-QCD calculation of the masses and decay constants of the positive-parity heavy-strange mesons $D^*_{s0}$, $D_{s1}$, $B^*_{s0}$,  and $B_{s1}$.
The calculations are performed with domain-wall fermions for the light and strange quarks and an anisotropic clover action for the charm and bottom quarks. We use seven different RBC/UKQCD ensembles with pion masses ranging from a near-physical 139 MeV up to 431 MeV. We consider two different analysis types, with or without two-meson operators at the source. We observe the expected below-threshold ground states. The fits without the two-meson operators appear to be more stable, but may overestimate the ground-state energies, while preliminary fits with two-meson operators at the source only appear to underestimate the ground-state energies.

\vspace{2ex}

\textcolor{red}{\bfseries This version includes an erratum with updated results after correcting an error in the data analysis code.}
}

\FullConference{The 41st International Symposium on Lattice Field Theory (LATTICE2024)\\
 28 July - 3 August 2024\\
Liverpool, UK\\}

\begin{document}
\maketitle

\section{Introduction}
There has been considerable interest in the positive-parity heavy-strange mesons since the 2003 discovery of the $D^*_{s0}(2317)^\pm$ by the BaBar collaboration \cite{implications,babar}. The $D^*_{s0}(2317)^\pm$ state lies $\sim 45 $ MeV below the $DK$ kinematic threshold, which is far away from the $n^{2s+1} \ell_J=1^3 P_0$  $q\bar{q}$ state expected from potential models. For example, \textit{Godfrey and Kokoski} predict a broad resonance above threshold \cite{quarkmodelexpectations}. This discrepancy prompts an interpretation of the $D^*_{s0}(2317)^\pm$ as an exotic state. Further possibly non-$q\bar{q}$ states are known from experiment, including the $J^P=1^+$ $D_{s1}(2460)^\pm$ sitting below and near the $D^* K$ threshold, as shown in Table \ref{tab:Ds}.

The $D^*_{s0}$ and $D_{s1}$ states have been identified as candidates for a molecular structure \cite{implications,guo}, that is, a four-quark state consisting of two shallowly bound $q\bar{q}$ color singlet pairs. This is based on key features such as nearness to the threshold and a relatively strong coupling to the meson-meson scattering-state decay channel \cite{barnes}. Weinberg's compositeness criterion is also applicable to assess the molecular structure of mesons \cite{guo, weinberg}. Lattice studies of the positive-parity $D_s$ spectrum date back to 1997 \cite{earlywork}, preceding the $D^*_{s0}(2317)^\pm$ discovery. These early studies failed to resolve the below-threshold state. After introducing interpolating fields of the form of two-meson pairs, a number of works have found a spectrum consistent with the BaBar measurements \cite{latticeDs, bali}. Further recent results for the $D_{s0}^*$ at heavier pion masses are reported in Refs.~\cite{Cheung:2020mql, Alexandrou:2019tmk}. 

In the bottom-strange sector, less is known about the positive-parity spectrum from experiment. The $B_{s1}(5830)$, as well as more massive states not yet assigned a spin, have been identified above threshold and are expected to be excited states. The ground states $B^*_{s0}$ and $B_{s1}$ have been studied on the lattice \cite{Burch:2008qx,lang,mohler}, and their proximity to the $B^{(*)}K$ thresholds also raises the possibility of molecular structure.

Another interesting feature of the heavy-strange scalar and axial vector mesons is their relationship with form factors for semileptonic decays, which may be used to probe new physics. Discrepancies between SM predictions and LHC measurements of branching fractions and angular observables have been observed for some processes involving semileptonic $b$ decays (the so-called ``$b$ anomalies'')  \cite{banomolies}. Currents for $b\rightarrow s$ (or $c\rightarrow s$) transitions have overlap with heavy-strange two-meson scattering pairs, and the form factors inherit the pole structure of the $B^{(*)} K$ ($D^{(*)} K$) amplitudes. Precise knowledge of the pole locations is helpful in parametrizing the form factors. Furthermore, when considering unitarity bounds on the form factors, the contributions from single-particle bound states in these amplitudes depend on their decay constants (see, e.g., Ref.~\cite{dispersivebounds}). The decay constants are known with high precision for the ground-state negative-parity heavy-strange mesons \cite{FlavourLatticeAveragingGroupFLAG:2024oxs}, but are less well known in the positive-parity case. For the positive-parity mesons, the decay constants are defined through $\bra{0} J_V^\mu \ket{H^*_{s0}}=i f_{H^*_{s0}} p^\mu$ and $\bra{0} J_A^\mu \ket{H_{s1}}=f_{H_{s1}} m_{H_{s1}} \epsilon^\mu$, where $J_V^\mu=\bar{Q}\gamma^\mu s$ and $J_A^\mu=\bar{Q}\gamma^\mu\gamma_5 s$.

The positive-parity charm-strange decay constants $f_{D^*_{s0}}$ and $f_{D_{s1}}$ have previously been calculated on the lattice \cite{bali}. For the bottom-strange decay constants $f_{B^*_{s0}}$ and $f_{B_{s1}}$, there appear to be no published lattice calculations, but there are estimates from QCD sum rules and other methods (see, e.g., Refs.~\cite{sr, Li:2018eqc}). In the following, we present preliminary results from a new lattice-QCD study of the positive-parity heavy-strange mesons, including, for the first time, the decay constants of the $B^*_{s0}$ and $B_{s1}$.

 \begin{table}
\centering
\begin{tabular}{ |c | c | c| c | } 
 \hline
 $n^{2s+1} \ell_J$  & $J^P$ & State & Mass\\
 \hline
 $1^1 S_0$ & $0^-$ & $D_s^\pm$ & 1968  MeV\\
 $1^3 S_1$ & $1^-$ &  $D_s^{*\pm}$ & 2112 MeV\\
 $1^3 P_0$ & $0^+$ & $D_{s0}^*(2317)^\pm$ & 2318 MeV\\
 $1^3 P_1$ & $1^+$ &  $D_{s1}(2460)^\pm$  & 2460 MeV\\
 $1^1 P_1$ & $1^+$ &  $D_{s1}(2536)^\pm$  & 2535 MeV\\
 \hline
\end{tabular}
\caption{Observed $D_s$ states \cite{pdg}. The $DK$ and $D^{*}K$ kinematic thresholds are located at $\sim 2.36$ GeV and $\sim 2.50$ GeV, respectively.}
\label{tab:Ds}
\end{table}

\section{Lattice Action and Ensemble Details}
Our simulations use Iwasaki-action gauge configurations with 2+1 dynamical domain-wall fermions produced by the RBC/UKQCD collaboration \cite{ukqcd, boyle}. Domain-wall fermions allow for exact chiral symmetry at finite lattice spacing where 4D chiral fermions are realized as low-energy degrees of freedom from a structure in a 5th dimension \cite{kaplan}. Of the 7 ensembles on which simulations are run (Table \ref{tab:ensembles}), C00078 and F1M use the M{\"o}bius domain-wall fermion formulation of \textit{Brower, Neff and Orginos} \cite{mobius}, and the others use the formulation of \textit{Shamir} \cite{shamir}. 

Table \ref{tab:ensembles} contains the lattice spacings, sizes, light and (partially quenched) strange-quark masses for each of our ensembles (see also \cite{Meinel:2023wyg}). Correlation functions are constructed via covariant approximation averaging \cite{AMA} over different source locations, using a total of 
$N_{\rm ex}$ and $N_{\rm sl}$ ``exact'' and ``sloppy'' samples on each ensemble. The sloppy samples were obtained using propagators with reduced conjugate-gradient iteration count, combined with low-mode deflation for the light quarks.

Ensemble C00078 has a near-physical pion mass, and all ensembles have large volumes, corresponding to $3.86 \lesssim m_\pi L \lesssim 6.09$. Note that C005LV is simply a larger-volume version of C005, which provides a convenient test of volume dependence.

\begin{table}[H]
    \centering
    {\footnotesize
\begin{tabular}{lccccccccc}
    \hline\hline
    Label & $N_s^3\times N_t $  & $a$ [\text{fm}]  & DW Type  &  $am_{u,d}$ &  $m_\pi$ [\text{GeV}] & $am_{s}^{(\mathrm{sea})}$ 
    & $am_{s}^{(\mathrm{val})}$  & $N_{\rm ex}$ & $N_{\rm sl}$ \\
    \hline
    C00078 & $48^3\times96$  & 0.114 & M{\"o}bius   & $0.00078$  & $0.13917(35)$  & $0.0362$  & $0.0362$  & 158  & 5056 \\
    C005LV & $32^3\times64$  & 0.111 & Shamir  & $0.005$    & $0.3398(12)$   & $0.04$    & $0.0323$  & 186 & 5022 \\
    C005   & $24^3\times64$  & 0.111 & Shamir  & $0.005$    & $0.3398(12)$   & $0.04$    & $0.0323$  & 311 & 9952 \\
    C01    & $24^3\times 64$ & 0.111 & Shamir  & $0.01$     &  $0.4312(13)$       & $0.04$    & $0.0323$  & 283 & 9056 \\
    F004   & $32^3\times64$  & 0.083 & Shamir  & $0.004$    & $0.3036(14)$   & $0.03$    & $0.0248$  & 251 & 8032 \\
    F006   & $32^3\times64$  & 0.083 & Shamir  & $0.006$    & $0.3607(16)$   & $0.03$    & $0.0248$  & 445 & 14240 \\
    F1M    & $48^3\times96$  & 0.073 & M{\"o}bius  & $0.002144$ & $0.2320(10)$   & $0.02144$ & $0.02217$ & 226 & 7232 \\
    \hline\hline
    \end{tabular}}
    \caption{Parameters for the ensembles and domain-wall propagators used in this work.}
    \label{tab:ensembles}
    \end{table}
    
The heavy quarks are handled with an anisotropic clover action tuned to remove discretization errors \cite{El-Khadra:1996wdx}, which allow us to work directly at the physical charm and bottom masses for all lattice spacings. Our action is of the form

\begin{equation*}
    S_Q=a^4\!\sum_x \bar{Q}\! \left[ m_Q + \gamma_0 \nabla_0 -\frac{a}{2} \nabla_0^{(2)} + \nu \!\sum_{i=1}^3  \left(\gamma_i \nabla_i - \frac{a}{2} \nabla_i^{(2)}\right) - c_E \frac{a}{2} \!\sum_{i=1}^3 \sigma_{0i} F_{0i} - c_B \frac{a}{4} \!\sum_{i,j=1}^3 \sigma_{ij} F_{ij} \right]\! Q.
\end{equation*}
The bare mass $am_Q$, anisotropy coefficient $\nu$, and clover coefficients $c_B=c_E$ are tuned by matching the $D_s^{(*)}$ and $B_s^{(*)}$ dispersion relations and hyperfine splittings \cite{Meinel:2023wyg, RBC:2012pds}. The values of these parameters are given in Ref.~\cite{Meinel:2023wyg} (which labels the ``C00078'' ensemble ``CP'').

The currents used to calculate the decay constants are renormalized according to the mostly-nonperturbative method of Refs.~\cite{Hashimoto:1999yp, El-Khadra:2001wco}, and have the form
\begin{align*}
    J_{V_0}&=\sqrt{Z_V^{ss}Z_V^{QQ}}\rho_{V_0}  \left[\bar{s} \gamma_0 Q + 2a \left(c^R_{V_0} \Bar{s} \gamma_0 \gamma_j \overrightarrow{\nabla}_j Q + c^L_{V_0} \Bar{s} \overleftarrow{\nabla}_j \gamma_0 \gamma_j  Q    \right)\right],\\
    J_{A_i}&=\sqrt{Z_V^{ss}Z_V^{QQ}}\rho_{A_i}  \Big[\bar{s} \gamma_i \gamma_5 Q \\ &\hspace{0.18\linewidth}+ 2a \left(c^R_{A_i} \Bar{s} \gamma_i \gamma_5 \gamma_j \overrightarrow{\nabla}_j Q + c^L_{A_i} \Bar{s} \overleftarrow{\nabla}_j \gamma_i \gamma_5 \gamma_j  Q  + d^R_{A_i} \bar{s} \gamma_5 \overrightarrow{\nabla_i} Q +  d^L_{A_i} \bar{s} \overleftarrow{\nabla}_i \gamma_5 Q \right)\Big].
\end{align*}
Here, the factors $Z^{qq}$ are computed nonperturbatively \cite{ukqcd,Boyle:2017jwu,Meinel:2016dqj,Meinel:2021rbm,Meinel:2023wyg}, while the residual matching factors $\rho_J$ and $\mathcal{O}(a)$-improvement terms are calculated to one loop in lattice perturbation theory \cite{detmold, lehner}.

\section{Operator Basis and Analysis}
\subsection{Operator Basis}

To resolve the low-lying heavy-strange states of interest and compute the decay constants, we use the operator basis
\begin{subequations}    
   \begin{gather}
    \Phi^{(1)}=\Bar{Q}s, \label{eq:spin0simple}\\
    \Phi^{(2)}=\Bar{Q}\gamma^i \nabla_i s, \\
    J_{V_0}=\sqrt{Z_V^{ss}Z_V^{bb}}\rho_{V_0}  \left[\bar{Q} \gamma_0 s + \mathcal{O}(a)\text{-terms}\right],    \\ 
    \Phi^{(4)}(\vec{x},t)=\sum_{\vec{y}} \Phi_K(\vec{x},t)\Phi_H(\vec{y},t) \ \text{  where } \Phi_K=\Bar{u}\gamma_5 s, \ \Phi_H=\Bar{Q}\gamma_5 u
    \end{gather}
\end{subequations}
for $J^P=0^+$, and 
\begin{subequations}
    \begin{gather}
    \Phi^{(1)i}=\bar{Q}\gamma^i\gamma_5 s,\\
    \Phi^{(2)i}=\bar{Q}\gamma_5 \nabla^i s,\\
    J_{A_i}=\sqrt{Z_V^{ss}Z_V^{bb}}\rho_{A_i}  \left[\bar{Q} \gamma_i \gamma_5 s + \mathcal{O}(a)\text{-terms}\right],\\
    \Phi^{(4)i}(\vec{x},t)=\sum_{\vec{y}} \Phi_K(\vec{x},t)\Phi_{H^*}(\vec{y},t) \ \text{ with } \ \Phi_K=\Bar{u}\gamma_5 s , \  \Phi_{H^*}=\bar{Q}\gamma^i u\label{eq:spin1mesonmeson}
    \end{gather}
\end{subequations}
for $J^P=1^+$. Inspecting Table \ref{tab:Ds} for the $J^P=1^+$ charm case, we expect the quark-model spin triplet $D_{s1}(2460)$ to couple strongly to $\Phi^{(1)i}$, and the quark-model spin singlet $D_{s1}(2536)$ to couple strongly to $\Phi^{(2)i}$ --- this will ultimately be what we find. For all systems, the two-meson operators $\Phi^{(4)(i)}$ are included to help resolve the $D^{(*)}K$ and $B^{(*)}K$ scattering states.

The light and strange quarks in all $\Phi$ fields are Gaussian-smeared using APE-smeared links, while the heavy quarks in the $\Phi$ fields are Gaussian-smeared using stout-smeared links. 

Using the above operators, we compute zero-momentum-projected correlation matrices 
\begin{equation}
    C^{lm}(t)=\sum_{\vec{z}} \langle \Phi^{(l)}(\vec{z}, t+t_s) \Phi^{(m)\dagger}(\vec{x}_s,t_s)\rangle,
\end{equation}
where $(\vec{x}_s,t_s)$ is the source position, and we use the convention that $\Phi^{(3)}$ corresponds to $J_{V^0}$ or $J_{A^i}$.

From previous projects \cite{Meinel:2016dqj,Meinel:2020owd,Meinel:2021rbm,Meinel:2023wyg} we have at our disposal precomputed light and strange quark propagators with Gaussian-smeared sources. To avoid having to compute new all-to-all light or strange propagators, we put the two-meson operators $\Phi^{(4)}$ only at the source. The elements $\sum_{\vec{z}}\langle \Phi^{(l)}(\vec{z}, t+t_s) \Phi^{(4)^\dagger}(\vec{x}_s,t) \rangle$ can then be computed using the light-quark propagator as a source for a sequential heavy-quark propagator, $T(z,x)\equiv \sum_{\vec{y}} G_Q(z;\vec{y},t_s) \gamma_5 G_u(\vec{y},t_s;\vec{x}_s,t_s) $. The correlation functions $\sum_{\vec{z}}\langle \Phi^{(l)}(\vec{z}, t+t_s) \Phi^{(2)^\dagger}(\vec{x}_s,t_s) \rangle$ with the derivative operators at the source are computed using heavy-quark propagators with derivative sources.

Where applicable, we average over Lorentz indices. We also average the off-diagonal $C_{lm}$'s with $C_{ml}$'s where both are available, as they have the same expectation values. Finally, we average over forward and backward propagation in time.

\subsection{Data Analysis and Fitting}\label{sec:data}
To determine the positive-parity spectra and decay constants, so far, we have focused on two different types of analyses. The first type combines a multi-exponential fit to the elements
\begin{equation}
\left( \begin{array}{llll} C_{11} & C_{12} & C_{13} & C_{14} \\ & C_{22} & C_{23} & C_{24} \\ &  & C_{33} & C_{34} \\
\end{array} \right),
\end{equation}
without the $O(a)$ improvement terms in the currents, with another multi-exponential fit in which no currents are included at the source and instead the $O(a)$-improvement terms on their own are included at the sink. The reason for this approach is that the implementation of the $O(a)$-improvement terms in $C_{l3}$ would require new strange-quark propagators with derivative sources.

The second type of analysis excludes the two-meson operators and uses single-exponential (for $J=0$) or two-exponential (for $J=1$) fits to the elements
\begin{equation}
\left( \begin{array}{ll} C_{11} &  \\
C_{21} & C_{22}  \\
C_{31} & C_{32}  \\
\end{array} \right),
\end{equation}
where the currents are included at the sink only and are fully $O(a)$-improved. The two-exponential fits are of the form $C_{ij}=A_i A_j\left(e^{-E t}+B_{1i} B_{1j} e^{-(E+\Delta E_1) t}\right)$ and are needed for $J=1$ because of the additional low-lying spin-singlet energy level expected from the quark model. An example fit of $C_{ij}$ and its associated effective-mass plot are exhibited below in Fig.~\ref{fig:corr}. The decay constants of the ground states are obtained through $f=A_3 \sqrt{\frac{2}{E}}$.

In addition to the above analyses of the positive-parity heavy-strange systems, we extract the masses of the $H^{(*)}=D^{(*)},B^{(*)}$ and $K$ mesons that determine the strong-decay thresholds from single-exponential fits to simple two-point functions using quark-antiquark interpolating fields.

\begin{figure}[H]
    \centering
    \includegraphics[width=0.99\linewidth]{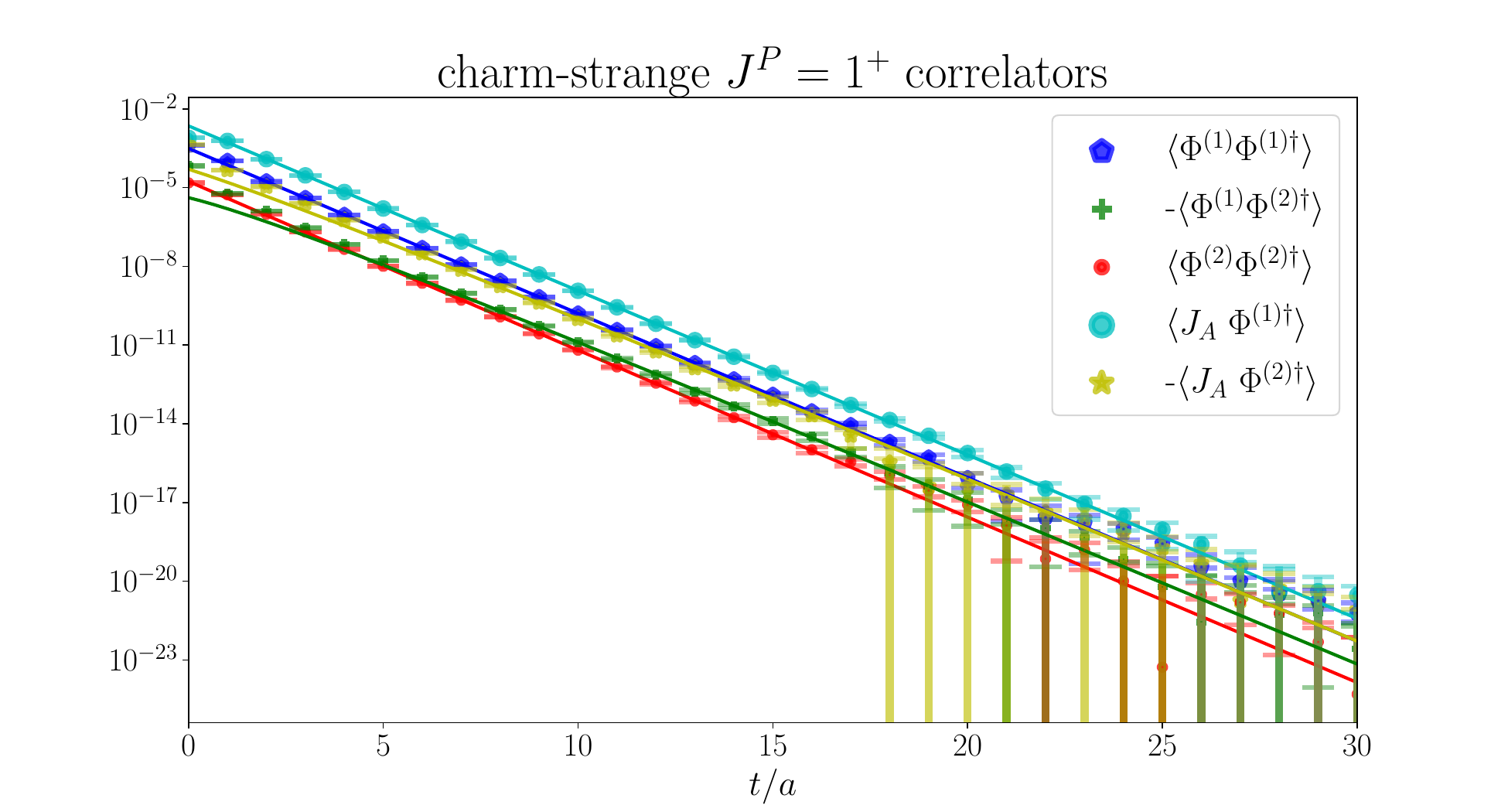}
    \hfill
    \includegraphics[width=0.99\linewidth]{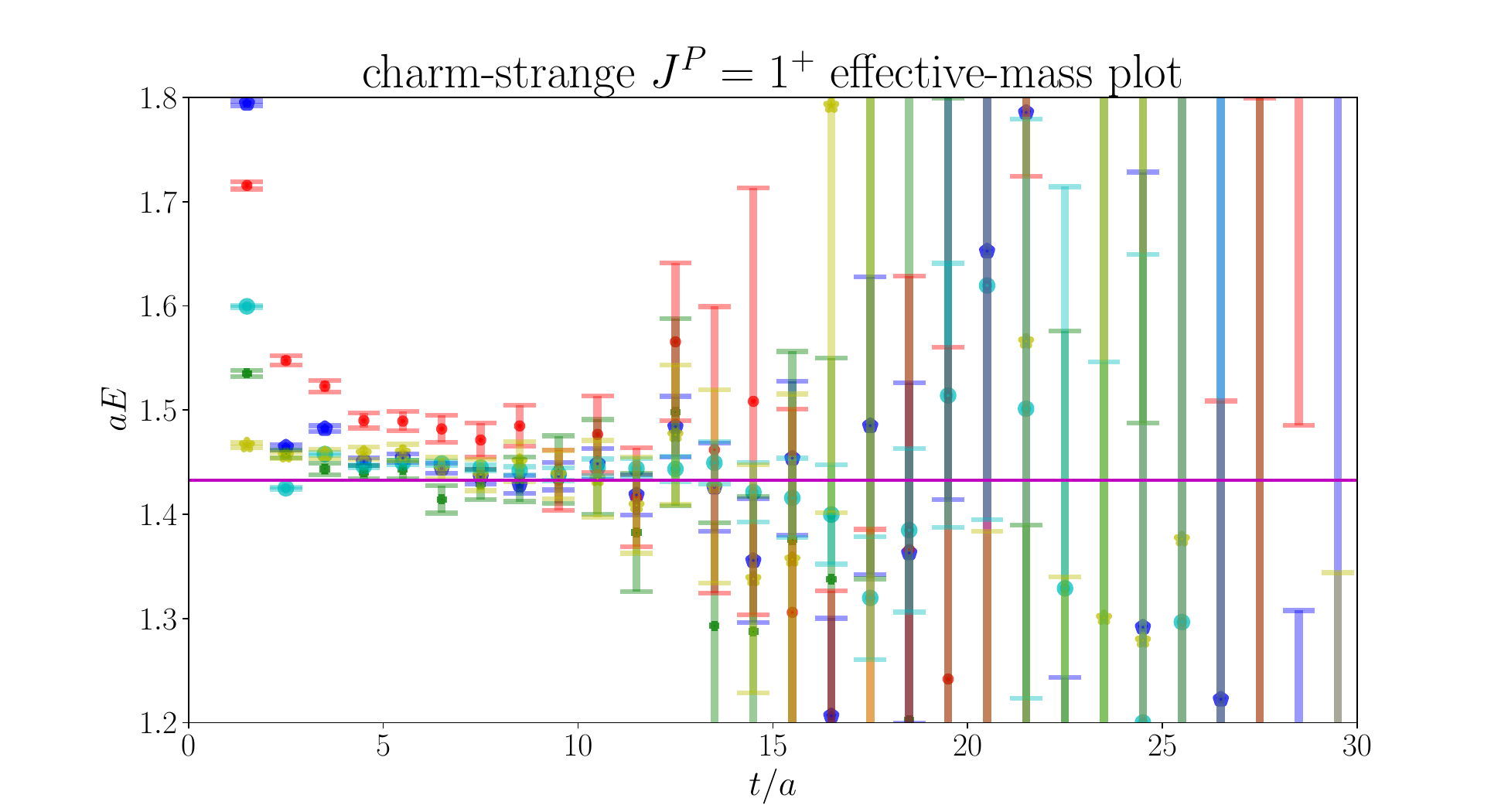}
    \caption{Correlation functions and corresponding effective-mass plot for the $D_{s1}$, fitted from $t_{\rm min}/a=10$ to $t_{\rm max}/a=16$ on the C00078 ensemble. The horizontal line shows the central value of fitted ground-state energy.}
    \label{fig:corr}
\end{figure}

\section{Results}\label{sec:results}

\textcolor{red}{\bfseries Our results in this section, which matches the version published in Proceedings of Science, are affected by an error in the code used for the data analysis. Please see the erratum at the end for the corrected results.}

\vspace{2ex}

The results presented at Lattice 2024 were obtained via the first type of analysis described in Sec.~\ref{sec:data}, with the two-meson operators  included at the source. These fits were extremely unstable and led to unexpectedly low ground-state energies. It has been observed that correlation functions with two-hadron operators at one end only and local operators at the other end can lead to a negative bias in the fitted energy due to the emergence of false plateaus \cite{Horz:2020zvv, Iritani:2016jie}. We have since performed new fits using the second type of analysis without the two-meson operators and found these fits to be substantially more stable and have lower values of $\chi^2/\text{d.o.f.}$. For this reason, we will only show the results from these new fits here.  However, concerns remain that without the two-meson operators there may now be an overestimation of the ground-state energy levels.

Our results for the finite-volume ``binding energies'' $\Delta_H=E-E_{H^{(*)}}-E_K$ and decay constants extracted from the second type of analysis are plotted below in Figs.~\ref{fig:Dsspec}-\ref{fig:Bsspec}. The error bars are calculated by combining, in quadrature, the statistical uncertainties with an estimate of the systematic uncertainty due to the choice of $t_\text{min}$, given by the shift in the central value when reducing $t_\text{min}/a$ by 1.

The results for the $B_{s0}^*$ and $B_{s1}$ decay constants exhibit a surprisingly strong dependence on the lattice spacing, but we emphasize that the fit results may still be unreliable due to the exclusion of the two-meson operators.

We plan to revisit the fits with the two-meson operators included, and we are considering combining results from both fit types. We further intend to employ Lüscher's method \cite{Luscher:1990ux} to extract infinite-volume bound state poles. Finally, we will perform chiral-continuum extrapolations of the results.

\begin{figure}[H]
\centering

\includegraphics[height=0.3\textwidth]{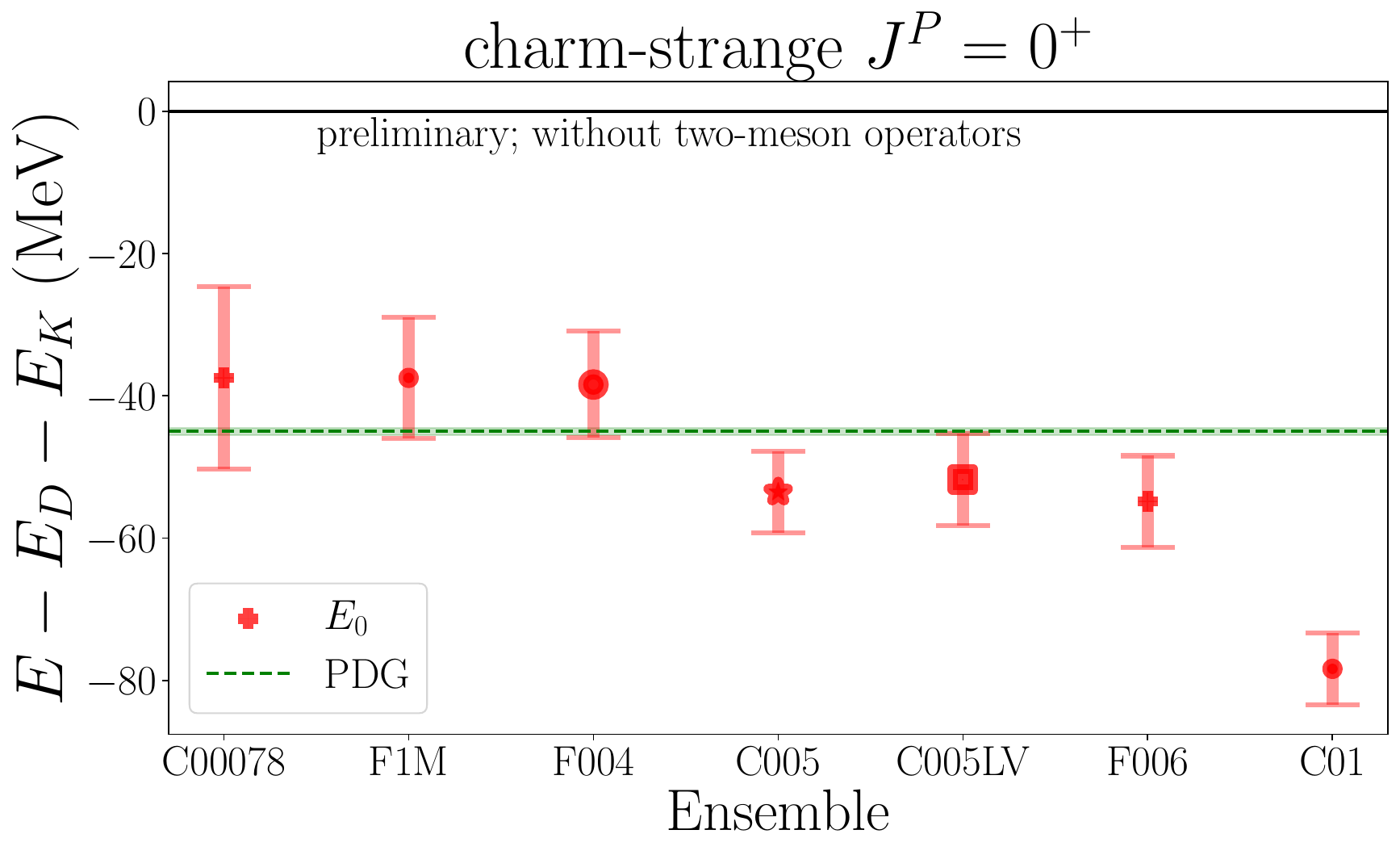}
\hfill
\includegraphics[height=0.3\textwidth]{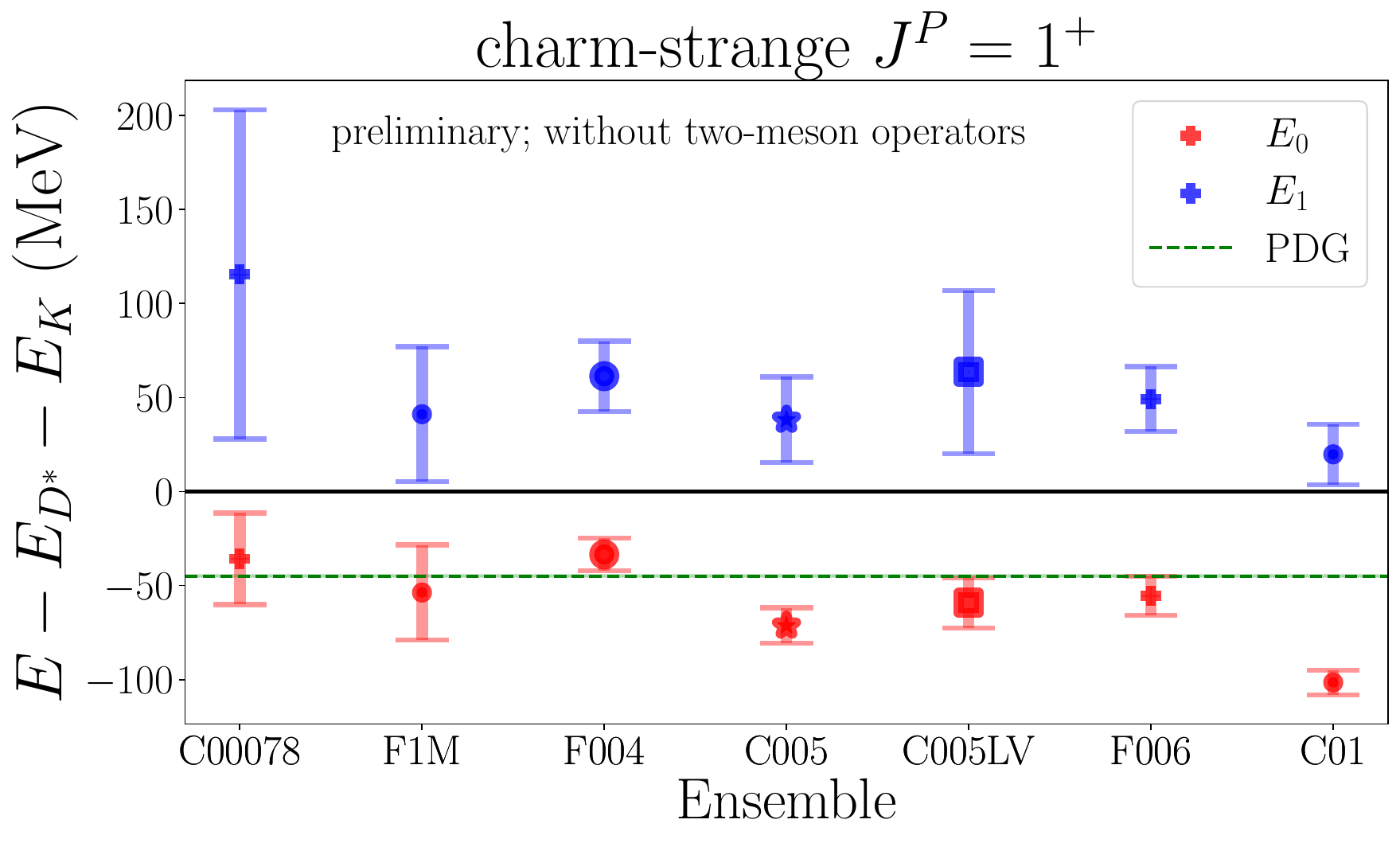}

\vspace{2ex}

\includegraphics[height=0.3\textwidth]{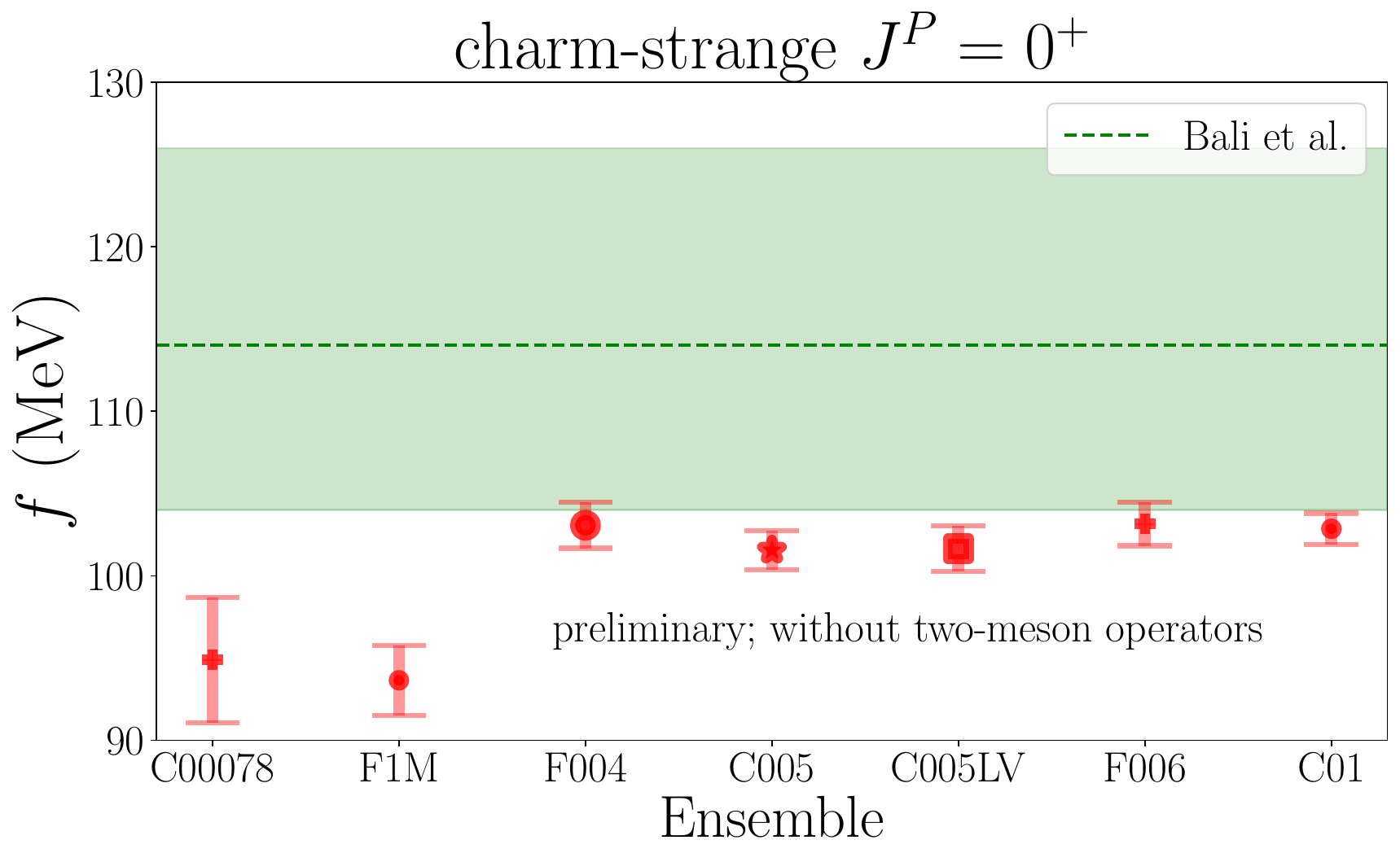}
\hfill
\includegraphics[height=0.3\textwidth]{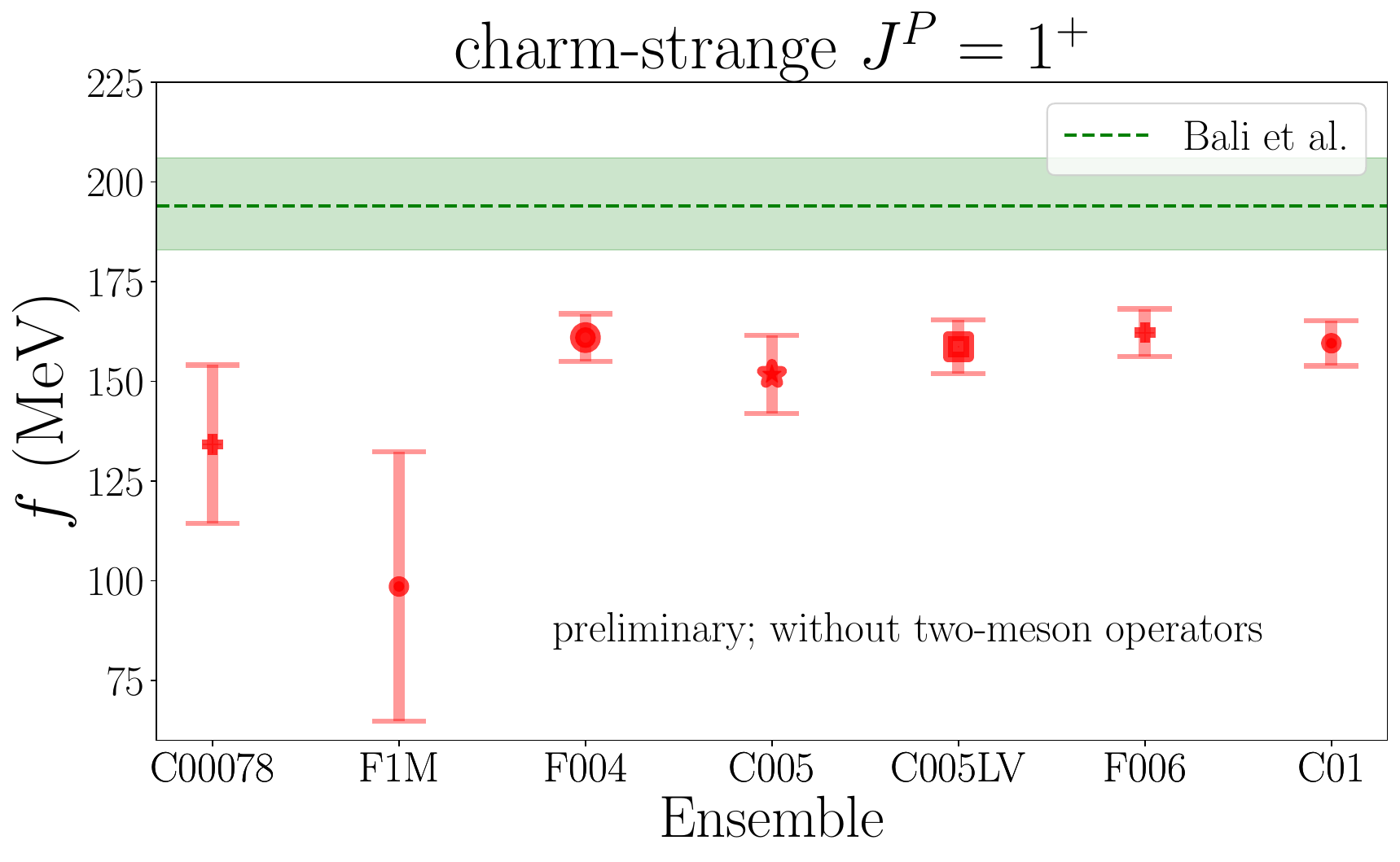}

\caption{Finite-volume spectrum and decay constants of the $D_{s0}^*$ and $D_{s1}$. For the spectrum, the bands show the experimental results for the ground state \cite{pdg}, and for the decay constants, the bands show the lattice results from Ref.~\cite{bali}.  \textcolor{red}{\bfseries Our results in this figure, which matches the version published in Proceedings of Science, are affected by an error in the code used for the data analysis. Please see the erratum at the end for the corrected results.}}
\label{fig:Dsspec}
\end{figure}

\begin{figure}[H]

\includegraphics[height=0.3\textwidth]{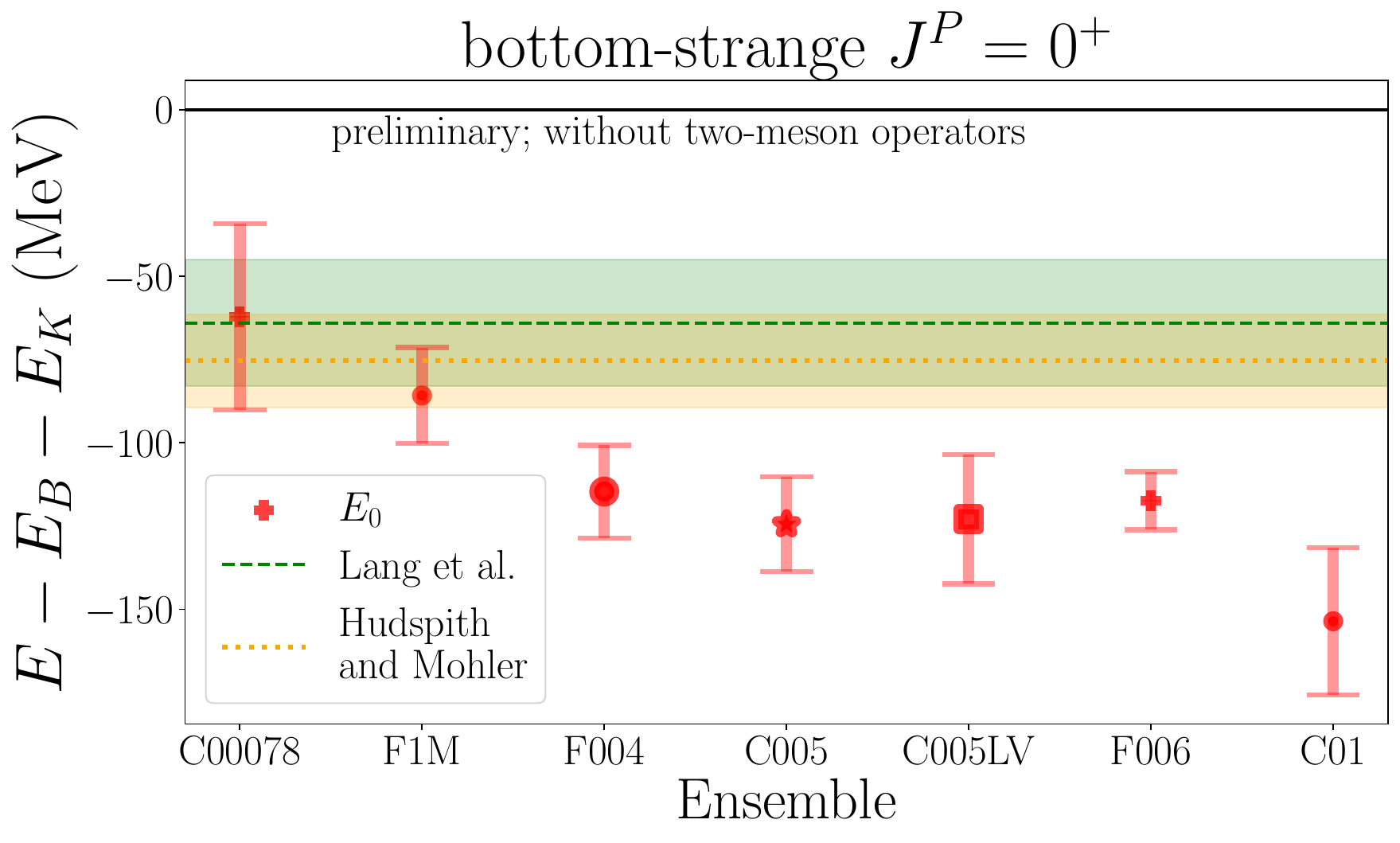}
\hfill
\includegraphics[height=0.3\textwidth]{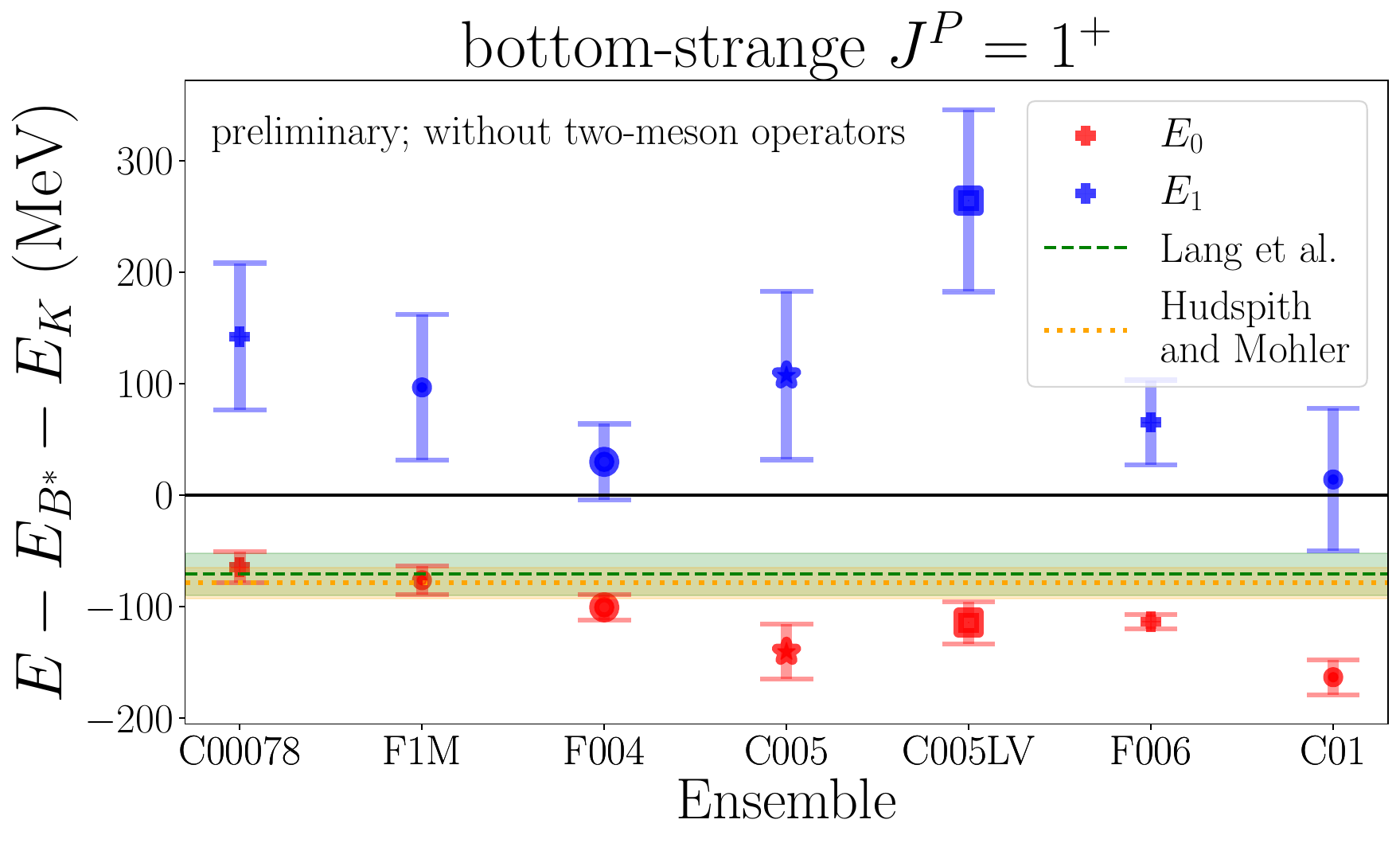}

\vspace{2ex}

\hfill \includegraphics[height=0.3\textwidth]{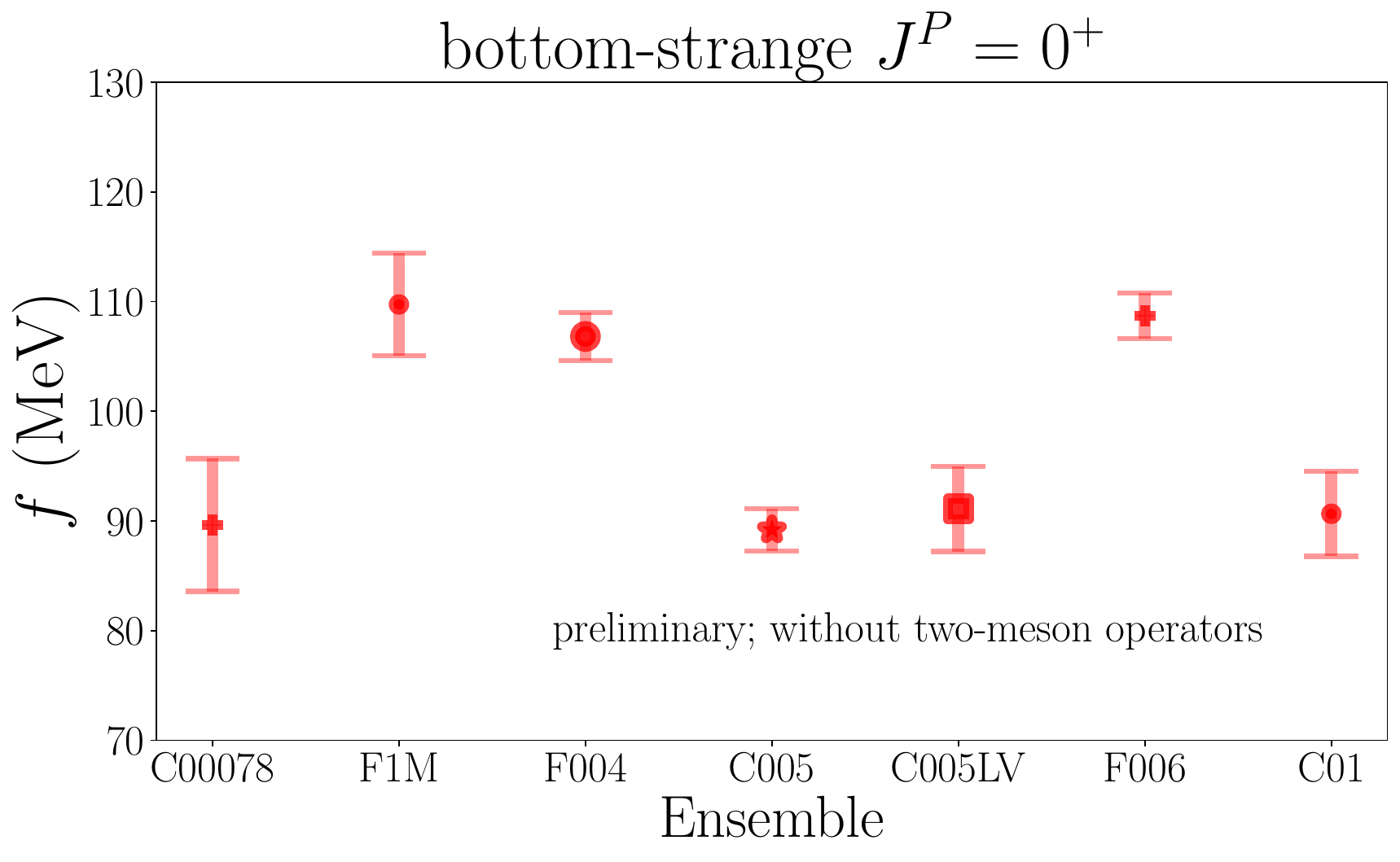}
\includegraphics[height=0.3\textwidth]{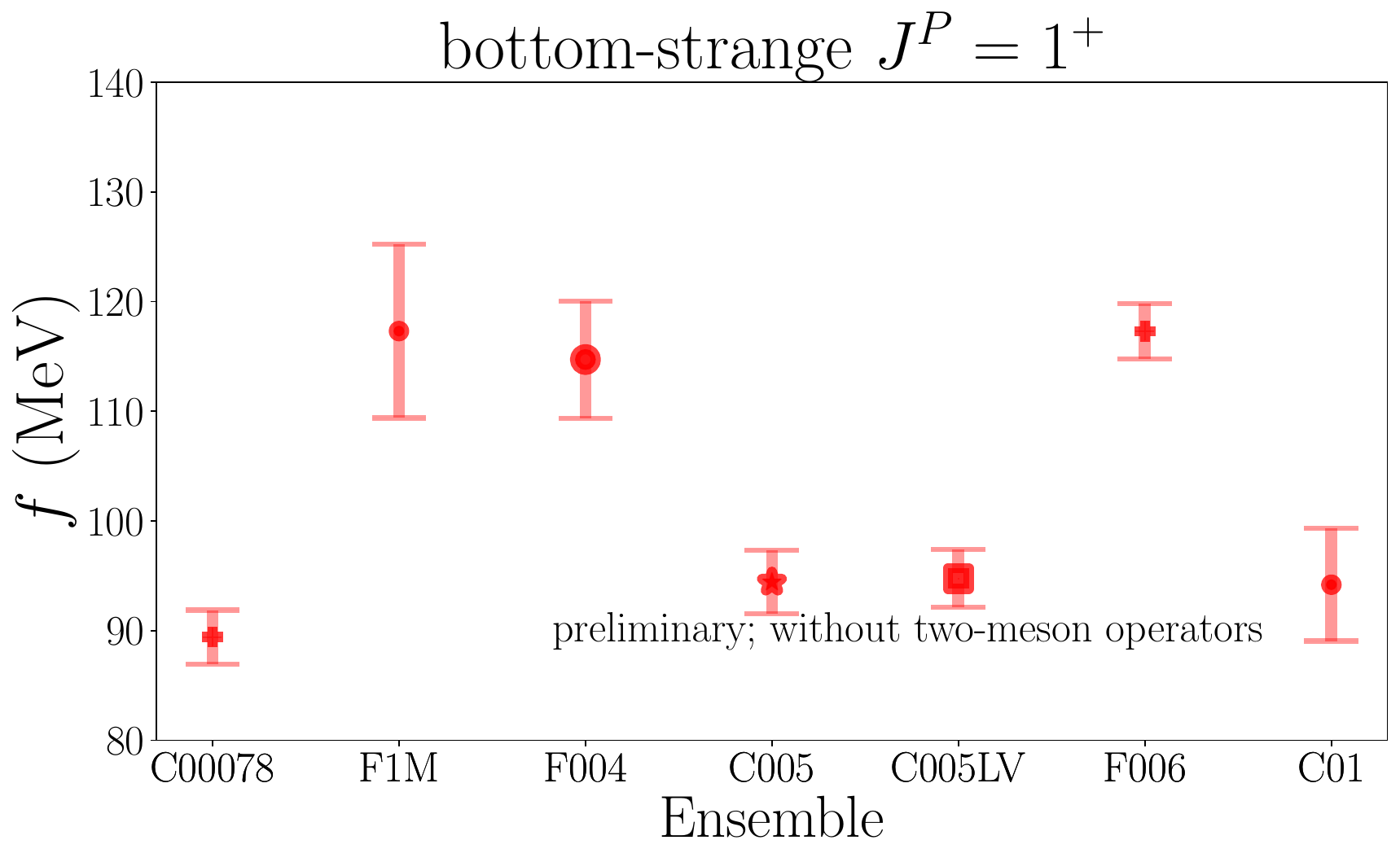}

\caption{Finite-volume spectrum and decay constants of the $B_{s0}^*$ and $B_{s1}$. Bands are infinite-volume estimates for the ground state from the lattice calculations of Refs.~\cite{lang,mohler}. \textcolor{red}{\bfseries Our results in this figure, which matches the version published in Proceedings of Science, are affected by an error in the code used for the data analysis. Please see the erratum at the end for the corrected results.}}
\label{fig:Bsspec}
\end{figure}

\section*{Acknowledgments}

We thank the RBC and UKQCD collaborations for providing the gauge configurations. We are supported by the U.S.\ Department of Energy, Office of Science, Office of High Energy Physics under Award Number DE-SC0009913. This work used resources of the National Energy Research Scientific Computing Center (NERSC), a U.S.\ Department of Energy Office of Science User Facility operated under Contract No.\ DE-AC02-05CH11231. This work also used the Anvil supercomputer at Purdue RCAC and the Ranch storage system at TACC through allocation PHY080005 from the Advanced Cyberinfrastructure Coordination Ecosystem: Services \& Support (ACCESS) program \cite{10.1145/3569951.3597559}, which is supported by U.S. National Science Foundation grants \#2138259, \#2138286, \#2138307, \#2137603, and \#2138296.

\section*{Erratum}

\begin{figure}
\includegraphics[height=0.3\textwidth]{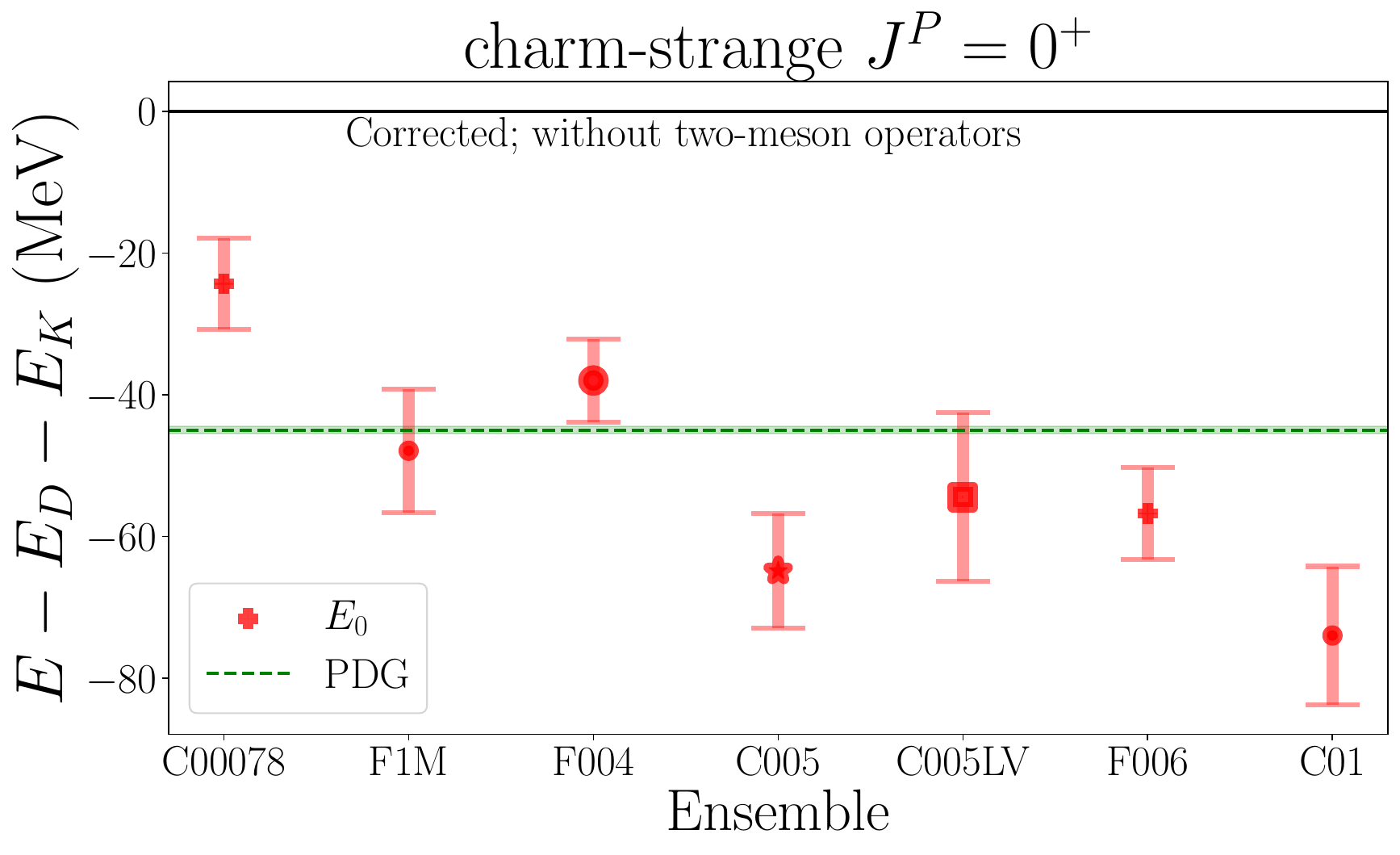}
\hfill
\includegraphics[height=0.3\textwidth]{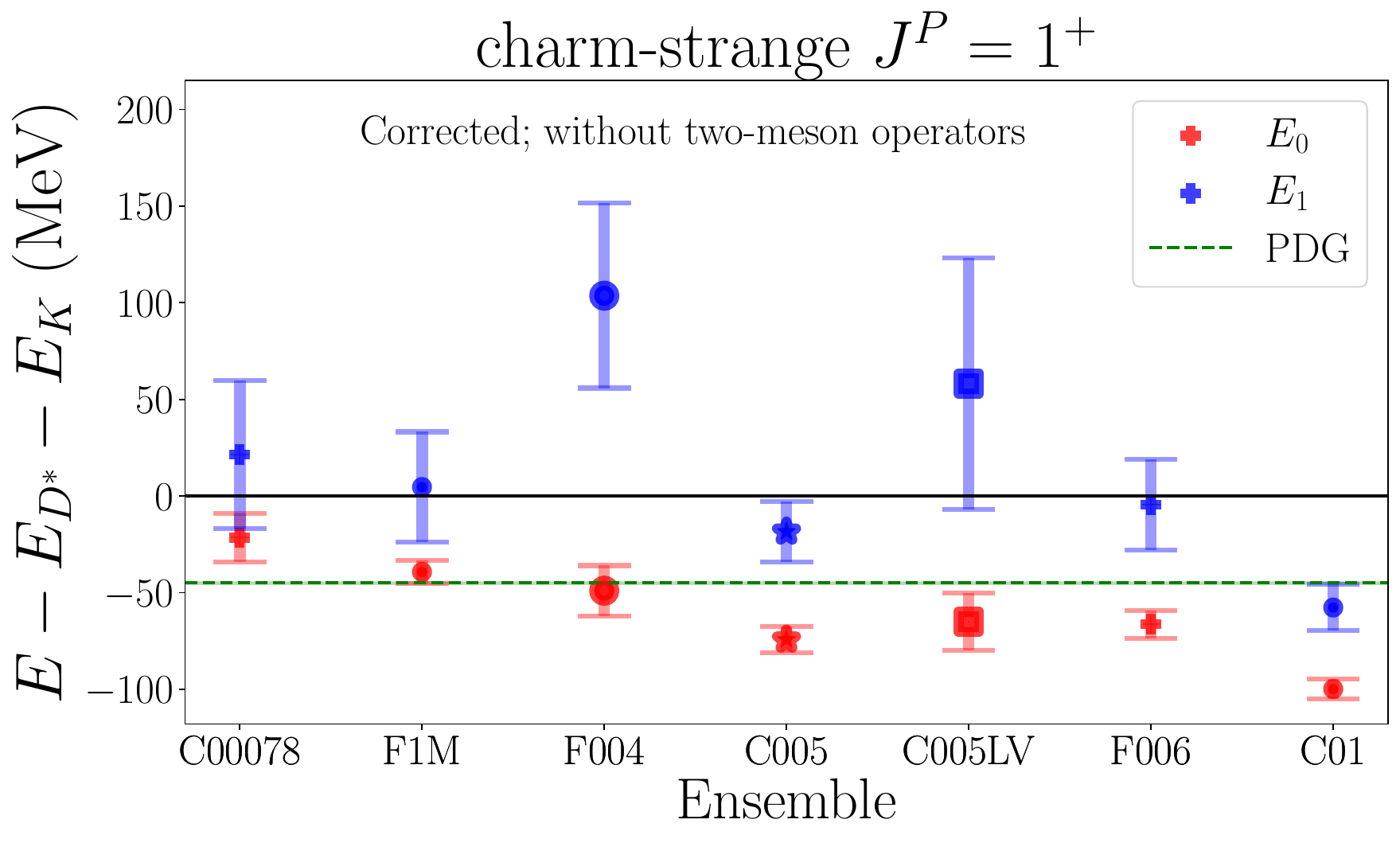}

\vspace{2ex}

\hfill \includegraphics[height=0.3\textwidth]{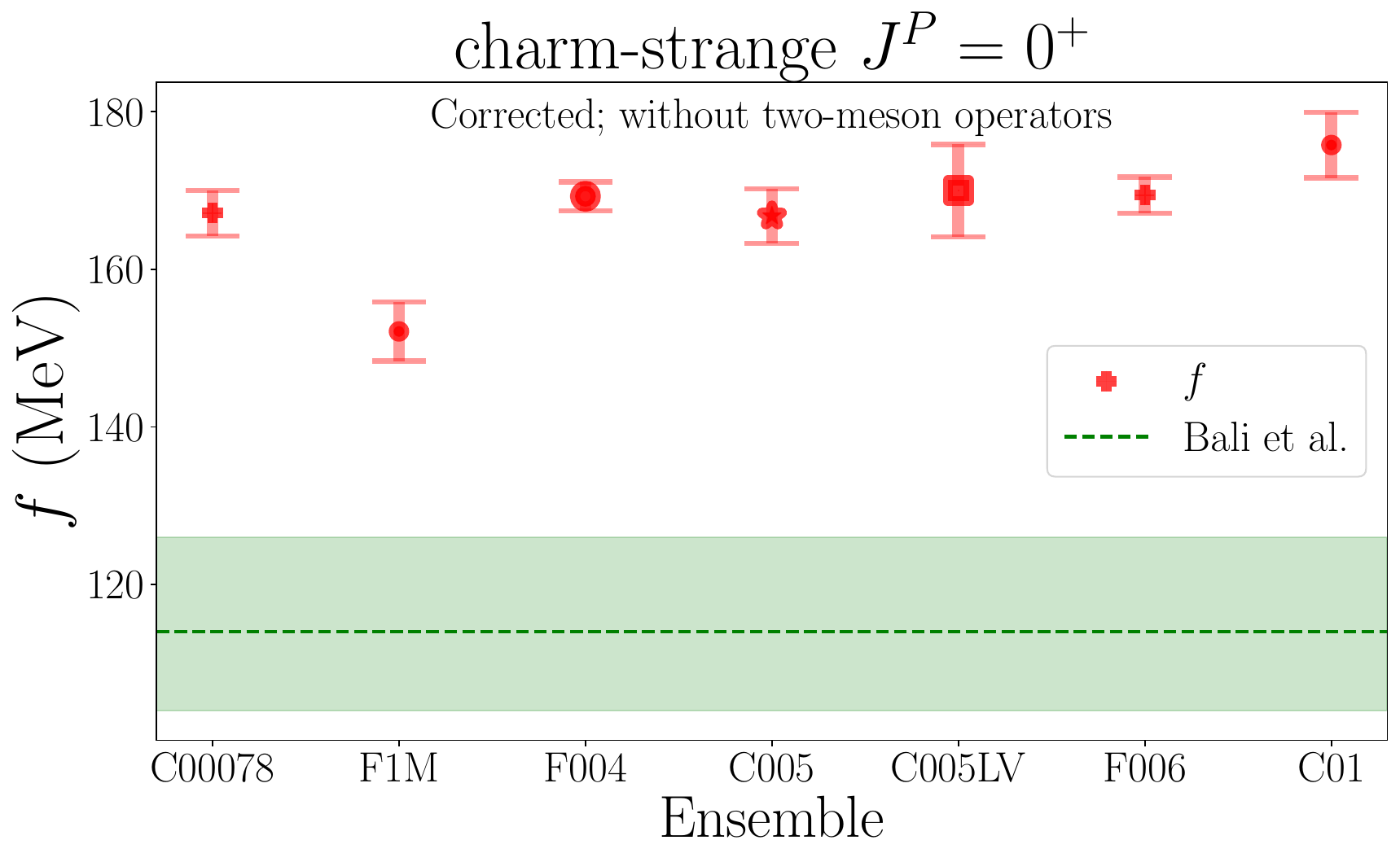}
\includegraphics[height=0.3\textwidth]{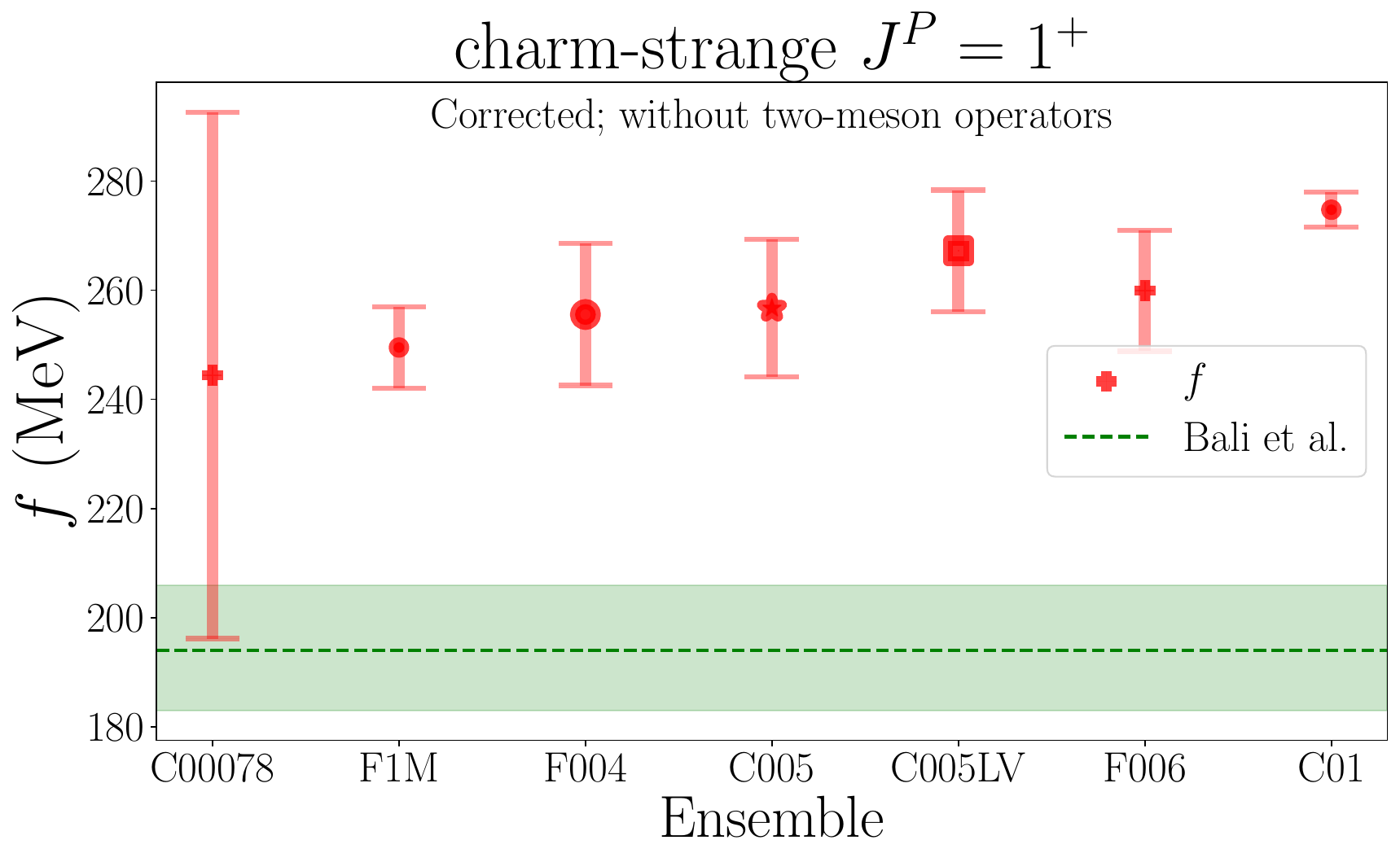}

\caption{Corrected finite-volume spectrum and decay constants of the $D_{s0}^*$ and $D_{s1}$. For the spectrum, the bands show the experimental results for the ground state \cite{pdg}, and for the decay constants, the bands show the lattice results from Ref.~\cite{bali}. }
\label{fig:Dsspec-corrected}
\end{figure}

\begin{figure}

\includegraphics[height=0.3\textwidth]{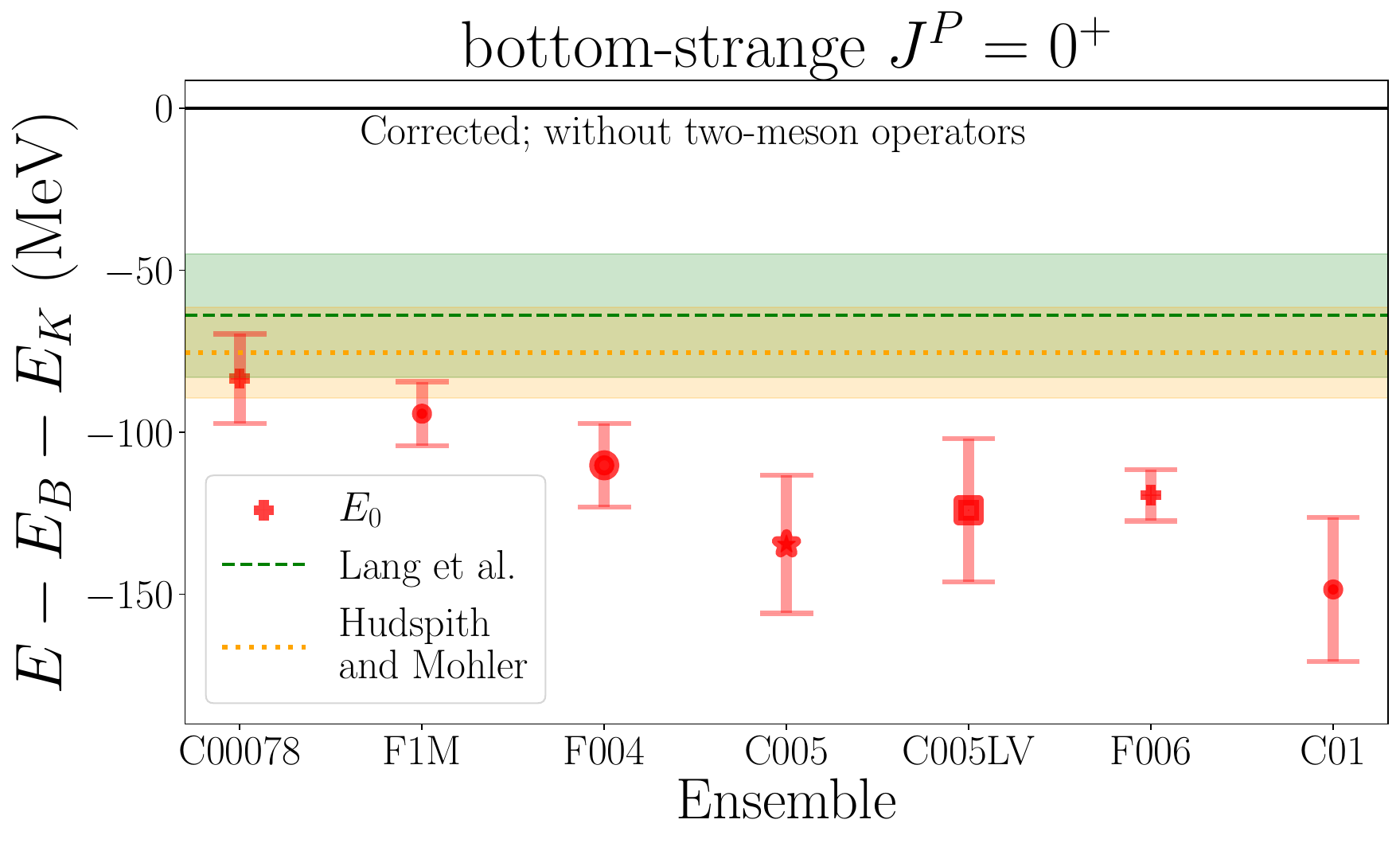}
\hfill
\includegraphics[height=0.3\textwidth]{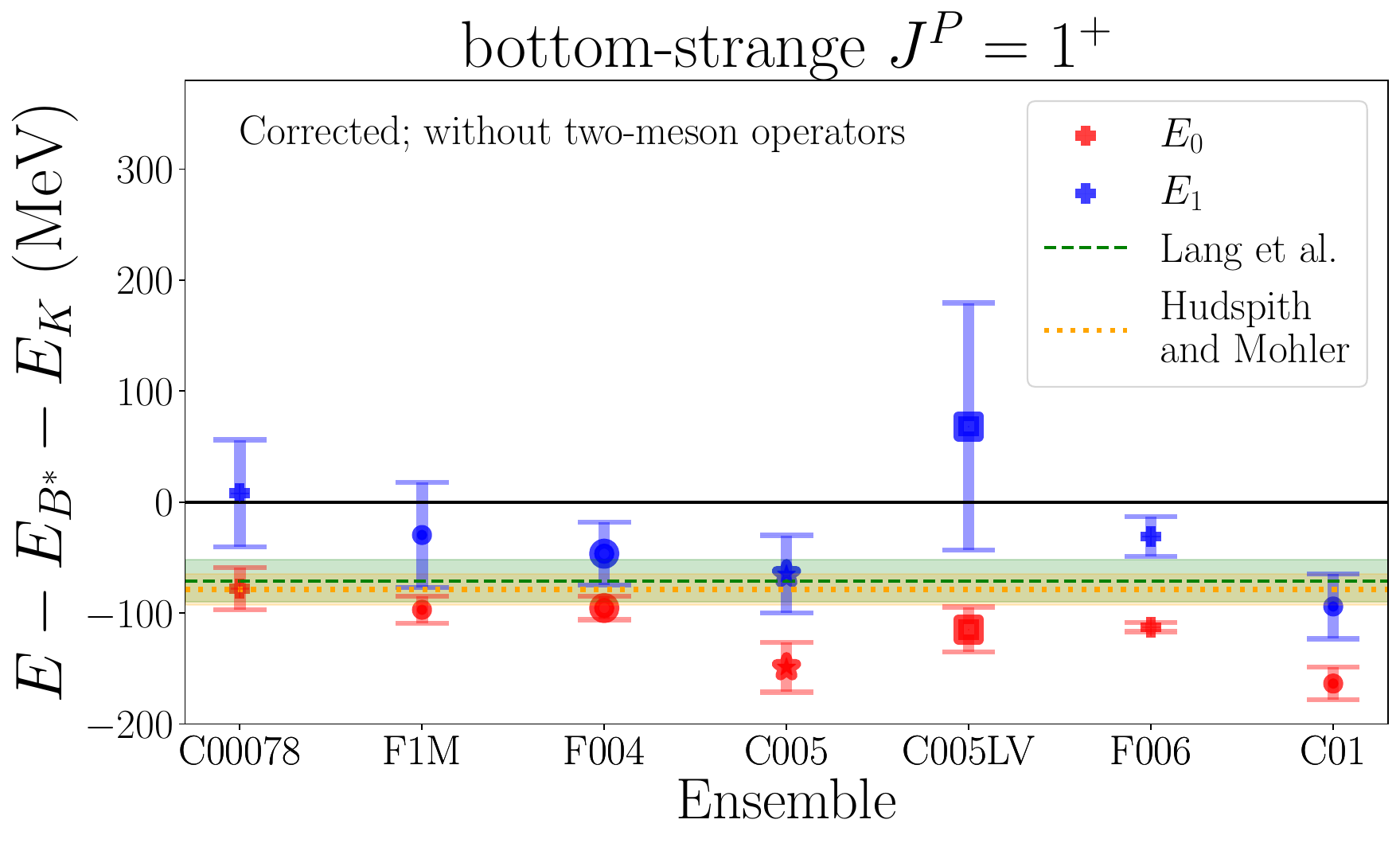}

\vspace{2ex}

\hfill \includegraphics[height=0.3\textwidth]{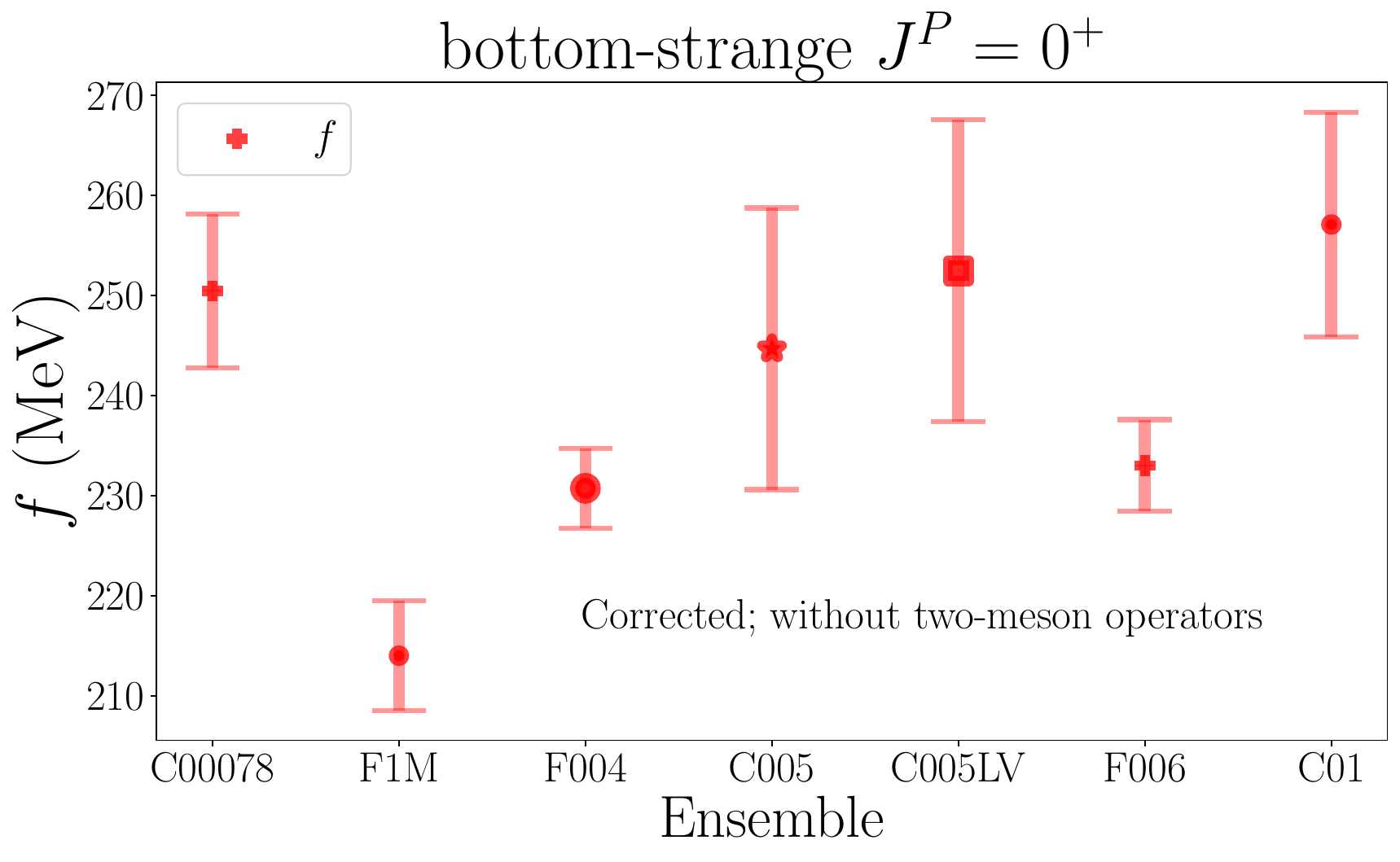}
\includegraphics[height=0.3\textwidth]{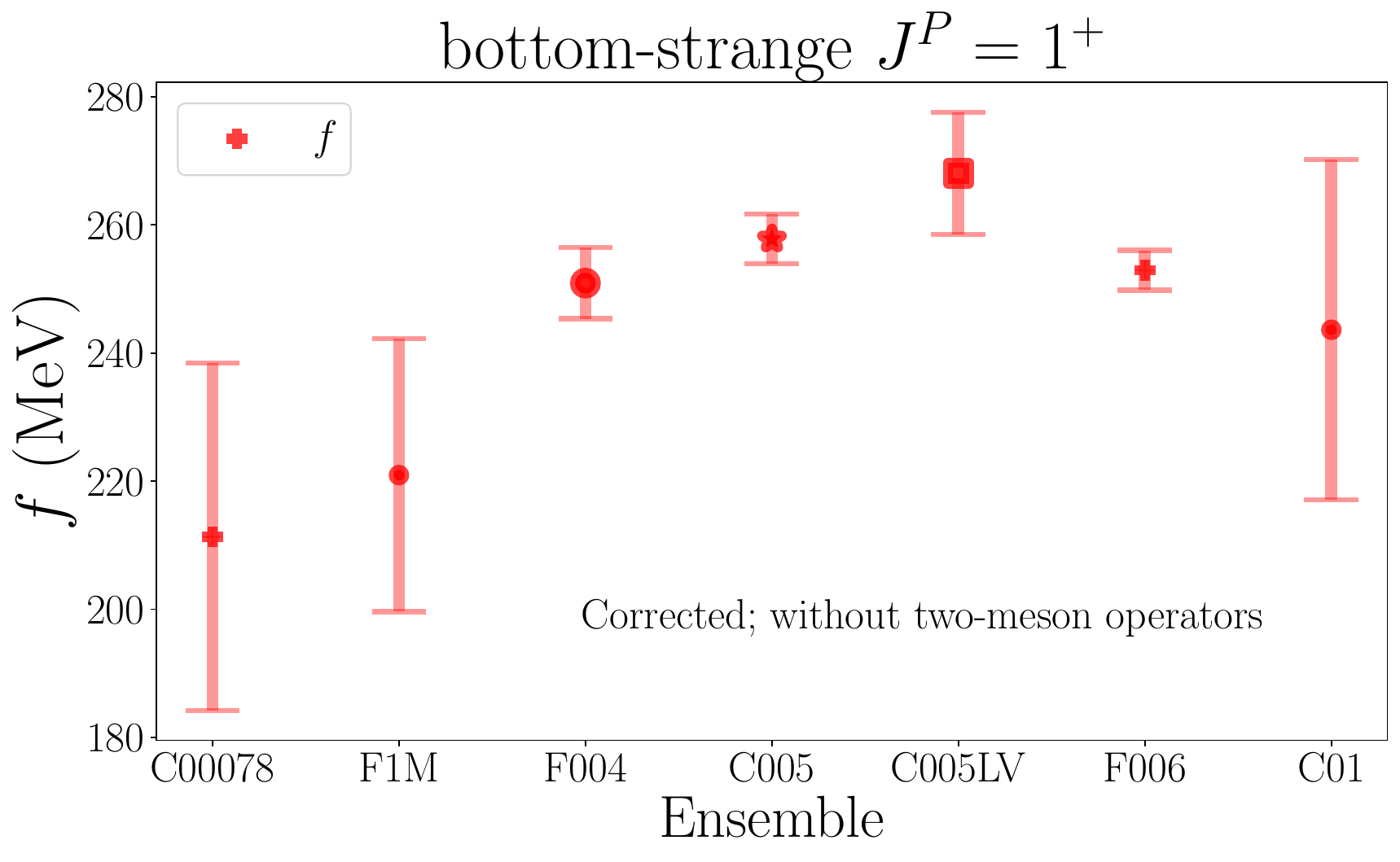}

\caption{Corrected results for the finite-volume spectrum and decay constants of the $B_{s0}^*$ and $B_{s1}$. Bands are infinite-volume estimates for the ground state from the lattice calculations of Refs.~\cite{lang,mohler}.}
\label{fig:Bsspec-corrected}
\end{figure}

The results shown above and in the Proceedings of Science version of this article are affected by an error in the code used for the data analysis (an indexing error in the averaging of the correlation functions over the forward and backward time directions). The corrected results are shown in Figs.~\ref{fig:Dsspec-corrected}-\ref{fig:Bsspec-corrected}. The values of the decay constants and their lattice spacing dependence have changed substantially.

\providecommand{\href}[2]{#2}\begingroup\raggedright\endgroup
\end{document}